# Ginzburg-Landau Like Theory for High Temperature Superconductivity in the Cuprates: Emergent d-wave Order*

T V Ramakrishnan
Department of Physics, Indian Institute of Science, Bangalore 560012, India
Department of Physics, Banaras Hindu University, Varanasi 221005, India

## Abstract

High temperature superconductivity in the cuprates remains one of the most widely investigated, constantly surprising, and poorly understood phenomena in physics. Here, we describe briefly a new phenomenological theory inspired by the celebrated description of superconductivity due to Ginzburg and Landau and believed to describe its essence. This posits a free energy functional for the superconductor in terms of a complex order parameter characterizing it. We propose that there is, for superconducting cuprates, a similar functional of the complex, in plane, nearest neighbor spin singlet bond (or Cooper) pair amplitude $\psi_{ij}$. Further, we suggest that a crucial part of it is a (short range) positive interaction between nearest neighbor bond pairs, of strength $J'$. Such an interaction leads to nonzero long wavelength phase stiffness or superconductive long range order, with the observed d-wave symmetry, below a temperature $T_c \sim z J'$ where z is the number of nearest neighbours; it is thus an emergent, collective consequence. Using the functional, we calculate a large range of properties, eg the pseudogap transition temperature $T^*$ as a function of hole doping x, the transition curve $T_c(x)$, the superfluid stiffness $\rho_s(x,T)$, the specific heat (without and with a magnetic field) due to the fluctuating pair degrees of freedom, and the zero temperature vortex structure. We find remarkable agreement with experiment. We also calculate the self energy of electrons hopping on the square cuprate lattice and coupled to electrons of nearly opposite momenta via inevitable long wavelength Cooper pair fluctuations formed of these electrons. The ensuing results for electron spectral density are successfully compared with recent experimental results for ARPES, and comprehensively explain strange features such as temperature dependent Fermi arcs above $T_c$ and the 'bending' of the superconducting gap below $T_c$.

# 1  Introduction

Superconductivity at unprecedentedly high temperatures was discovered[1] in 1986 in the doped cuprates (the original compound is $La_{2-x}Ba_xCuO_4$, which has x holes per formula unit). Since then, this family of compounds has been the subject of a very large number of investigations which have brought out in great detail their unusual electronic nature. Despite this activity over a generation, involving nearly two hundred thousand papers as well as many theoretical models, it is probably true that there is no comprehensive theoretical description of their 'normal' or non-superconducting state together with the origin of superconductivity in them. In this talk, after a very brief account of some properties of cuprates in this section, I motivate and describe a novel phenomenological Ginzburg Landau like theory of the (Cooper) pair degrees of freedom in the cuprates (Section II). The original Ginzburg Landau approach was proposed a few years before the microscopic Bardeen Cooper Schrieffer (BCS) theory of superconductivity and was connected with the BCS theory by Gor'kov very soon after the advent of the BCS theory. It is widely believed to capture the essence of superconductivity and is a development of the idea that the macroscopic coherence of a superconductor is described by a complex order parameter function $\psi(\mathbf{r}) = \Delta(\mathbf{r}) \exp(i\Phi(\mathbf{r}))$. It has been applied successfully to a bewildering range of phenomena in superconductors, especially those involving inhomogeneities. Our theory with a superficially similar functional of $\Delta_m$ and $\Phi_m$ (where m is a square lattice site) is characterized (among other things) by a pseudogap crossover temperature $T^*(x)$ around which the thermal average $<\Delta_m>$ becomes substantial without any long range superconducting order, and an emergent d-wave symmetry superconducting transition at $T_c(x)$. In Section III, I describe these consequences as well as others, such as the superfluid density $\rho_s(x,T)$, the specific heat $C_v(x,T)$ without and with a magnetic field, and the zero temperature vortex structure. This broad range of properties is seen to be in good agreement with experiments. Measurements mostly involve particles coupled to the Cooper pair (of electrons). Of these, the most common and precise are those that involve electrons. In order to address such observations, a minimal model in which electrons hopping on the square copper lattice admix quantum mechanically with pairs is then described (Section IV). The electron self energy naturally arising from such a coupling leads to the spectral function which is used to make qualitative and quantitative sense of unexpected high resolution ARPES results such as Fermi arcs above $T_c$ and 'bending' of the gap below $T_c$. In the concluding Section (Section V), I allude to quantum phase fluctuation effects, other degrees of freedom ignored in the present work, microscopic theory, etc.. Some of this work is published[2,3]; a few more papers are in preparation[4-6].

About a quarter of a century ago, superconductivity was discovered[1] in $La_{2-x}Ba_xCuO_4$. Since then, about thirty chemically distinct families of cuprate superconductors have been discovered with $T_c$ ranging upto 160K. They all consist of one or more closely spaced layer sets with each planar layer containing Copper and Oxygen atoms arranged in a square array, and with relatively weak coupling between these sets. Figure 1 shows two of these structures. The crucial low energy degree of freedom is provided by electrons in the $d^9$ configuration of the planar Copper ions. We will therefore assume here (as is commonly done) that superconductivity, due the coherent pairing of these electrons, primarily resides in the planes.

An idealized, universal, phase diagram of the cuprates in the doping, temperature or (x,T) plane is shown in Fig.2. These materials have unusual electronic properties, especially in the 'normal' metallic state above $T_c$ (for a relatively recent review, see ref.7). For example, there is a poorly understood regime characterized by a pseudogap[8] against electronic excitations (Fig.2). The prefix 'pseudo' is used because while the spectral density for electrons at the Fermi surface peaks at nonzero excitation energy, implying a gap in the one electron excitation spectrum, there is also nonvanishing strength at zero excitation energy unlike that for a real gap. The pseudogap region appearing below a crossover temperature scale $T^*$ is characterized by a host of physical changes, eg diminution of paramagnetic spin susceptibility and dip in the optical conductivity. There is also an ubiquitous 'strange metal' phase in which the electrical resistivity increases linearly with temperature[9] from low temperatures ($\geq T_c$ which can be as low as 10K) to very high temperatures of the order of several hundred degrees Kelvin. In the pseudogap phase (namely below $T^*$ and above $T_c$), ARPES experiments do not show a Fermi surface (expected for a normal metal, namely one without superconducting or any other kind of long range order) but instead exhibit Fermi arcs[10]. Namely, the zero excitation energy state in **k** space (the Fermi surface) in the Cu –O plane does not extend from the node (namely $k_x = \pm k_y$) to the Brillouin zone edge ( $|k_x| = |k_y| = (\pi/a)$ ) but only till a certain distance on the Fermi surface. Fig.3 shows schematically such Fermi arc observations. The Fermi arc increases in length with temperature, till at about $T^*$, it touches the Brillouin zone edge; that would be the conventional Fermi surface.

The measured ARPES spectral functions are very broad[11]. The ones near the zone edge (antinode) do not seem to have a quasiparticle peak; crudely this means that there are no electronic states with well defined momentum and energy near the Fermi surface! Below $T_c$, there is a quasiparticle peak for all **k**; in the antinodal region for example its strength increases with the superfluid density eg as T decreases towards zero[12]. Early ARPES observations[13] of the low temperature gap as a function of **k** on the Fermi surface helped establish a d-wave symmetry behavior for the gap , namely showed that $\Delta_\mathbf{k} = \Delta_o\{\cos(k_xa)-\cos(k_ya)\}$. In the last few years, many authors have shown using ARPES[14] as well as STS that as a function of **k** on the Fermi surface, the gap below $T_c$ is not exactly of the above form, but shows 'bending' with respect to it, the bending becoming more pronounced as temperature increases. This has been attributed to the generic existence of two gaps[15], one primarily nodal and another, antinodal. We find, however, as described in Ref.3 and in Section IV of this paper, that there is only one gap, and that the observed bending arises from the coupling of the gapped electronic quasiparticles to collective excitations of the d-wave symmetry pair field.

## II. Ginzburg Landau like theory

About forty years after the discovery of superconductivity by Kamerlingh Onnes in 1911, Ginzburg and Landau[16] represented the macroscopic coherence in a superconductor by proposing that the entire superconductor is characterized by a single complex 'wave function' $\psi(\mathbf{r}) = \Delta(\mathbf{r}) \exp(i\Phi(\mathbf{r}))$ where $\Delta$ and $\Phi$ are real, and further that its free energy F is a functional of $\psi(\mathbf{r})$. They used the form

$$F(\{\psi(\mathbf{r})\}) = \int [\, a_{cont}|\psi(\mathbf{r})|^2 + b_{cont}|\psi(\mathbf{r})|^4 + c_{cont}\,|\nabla\psi(\mathbf{r})|^2\,]\,. \qquad (1)$$

This form is appropriate near the transition temperature where $\psi(\mathbf{r})$ is small, so that in a power series expansion, it is sufficient to retain the lowest order terms. The odd power terms vanish because of time reversal invariance as well as the invariance of the free energy F with respect to global phase chage, ie the transformation $\Phi \to \Phi + \Phi_o$. The quantity $\psi(\mathbf{r})$ is coarse grained. Its most likely configuration satisfies the extremum condition $(\partial F/\partial \psi) = 0$; the corresponding equation for $\psi(\mathbf{r})$ is superficially like a nonlinear Schrodinger equation. Since inhomogeneities in $\psi(\mathbf{r})$ are expected to cost energy, $c_{cont}$ is taken to be positive. For the homogeneous case, $\psi(\mathbf{r}) = \psi$ independent of $\mathbf{r}$. For positive $b_{cont}$ (necessary on grounds of energetic stability) the free energy is a minimum (ignoring the contribution of spatial fluctuations in $\psi(\mathbf{r})$ to the free energy) for zero $\psi$ when $a_{cont}$ is positive, and for nonzero values of $\psi$ when $a_{cont}$ is negative. Thus, the change in sign of $a_{cont}$ from positive to negative marks the superconducting temperature $T_c$ (or broken symmetry temperature) in such a mean field theory. Very soon after the advent of the BCS[17] theory of superconductivity, Gor'kov[18] obtained the functional Eq.(1) from microscopic theory by eliminating the electronic degrees of freedom for a given $\psi(\mathbf{r})$, and identified it as the thermodynamically averaged local Cooper pair amplitude, namely $\psi(\mathbf{r}) = \langle a_\uparrow(\mathbf{r}) a_\downarrow(\mathbf{r}) \rangle$. The continuum coefficients $a_{cont}$, $b_{cont}$ and $c_{cont}$ were also obtained in terms of the parameters of the (free electron) metal. Generalizations of Eq.(1) for the situation where $\psi(\mathbf{r})$, the order parameter function, depends on time also exist. As is well known[19], the Ginzburg Landau theory is extremely successful for superconductors under a wide variety of situations (eg inhomogeneities), and is believed to capture the essence of superconductivity.

We propose here an analogous approach for cuprates. Most of the effort in the last generation has been to describe the phenomena in microscopic terms. However, their strangeness, which is most probably due to the fact that these systems are strongly correlated metals, remains a barrier to reliable understanding. We do not yet have a new paradigm for strongly correlated metals comparable to the Drude free electron gas theory of metals and its highly sophisticated developments (eg the Landau Fermi liquid theory). It is not clear that such a paradigm is possible, nor that it is a necessary starting point. For the cuprates themselves, there have been many microscopic theory based attempts (eg the strong correlation approaches such as the resonating valence bond (RVB) theory[20] and approaches using dynamical mean field theory or DMFT[20]) at explaining the nature and occurrence of superconductivity in them as well as consequences for normal state properties. These do not appear to be able to do justice to the range of observations. Given this, it is essential to have (even at this late date!) a phenomenological theory in the spirit of Ginzburg and Landau for the cuprates with changes appropriate for the differences from conventional or BCS superconductors. This could be a base theory for superconductivity in them, enabling one to rationalize a large range of phenomena, and could be expanded to encompass several competing 'phases'. Since there is direct evidence for Cooper pairing in the cuprates, pairs ought to be the basic low energy degrees of freedom associated with superconductivity (like $\psi(\mathbf{r})$ which is identified with the local Cooper pair amplitude in 'conventional' superconductors). How the existence of pairs translates into what is observed in the cuprates, eg d-wave symmetry superconductivity, the pseudogap temperature $T^*$ and the Fermi arc, is not obvious. We believe that our approach provides a comprehensive answer. In an intensely explored field like cuprate superconductivity, many of our ideas must have been proposed in some form. We do not attempt to completely connect with these.

We start with the square planar copper lattice (Fig. 4), in which the relevant electrons are at the lattice sites (a fraction (1-x) of the sites are occupied by electrons, and a fraction x are unoccupied). The electrons pair, and the pairs as well as the unpaired electrons hop quantum mechanically from site to site. We describe the resulting free energy functional as follows. The basic Cooper pair is taken to be the nearest neighbor spin singlet, i.e. the spin singlet arising out of the pairing of two electrons at nearest neighbor sites i and j on the square lattice (Fig.4). Such a pair can arise because of the nearest neighbour superexchange $J_{ij}$ which is known, from inelastic neutron scattering experiments, to be large and positive for at least the parent compound ($J_{ij}\sim$1500K). For spin (1/2) particles this is known identically to be the nearest neighbour pair attraction[21] with the same strength, i.e. it amounts to a strong nearest neighbour attraction. Since this is a short range interaction, its strength is expected to depend only weakly on hole doping x (eg in the well known Gutzwiller approximation for strong coupling effects, the factor multiplying $J_{ij}$ is $(1+x)^{-2}$ instead of 1). We assume therefore that the low energy degrees of freedom are, at temperatures and energies of interest to us, Cooper pairs formed because of this attraction[22]. These are characterized by a complex amplitude $\psi_{ij} = \Delta_{ij} \exp(i\Phi_{ij})$. For convenience, we denote the pair (ij) by the bond centre site m (Fig.4), so that our basic degrees of freedom are $\psi_m$. The sites m form another square lattice rotated by 45° with respect to the original. The free energy F of the superconductor is written, in the image of the original Ginzburg Landau theory, as a power series in $\psi_m$. The first two terms are proportional to $|\psi_m|^2$ and $|\psi_m|^4$, with coefficients a and b which are somewhat different from their continuum values $a_{cont}$ and $b_{cont}$ ( see below).

The third term which couples $\psi_m$ with the nearest neighbour pair amplitude $\psi_n$ ( Fig.4) is analogous to the gradient term in the continuum. For a given pair (m,n) this term has two extrema, one at $(\Phi_m - \Phi_n) = \Phi = 0$ and another at $\Phi = \pi$. This extremum is a free energy minimum for negative c(x,T) in the former case, and for positive c(x,T) in the latter case. The continuum limit of the former is obvious; for m and n infinitesimally removed from each other, $(\Phi_m - \Phi_n)$ is very small so that one can make a Taylor series expansion in powers of the distance (m-n) and the free energy cost of small inhomogeneities in the phase translates into a gradient term as present in the original Ginzburg Landau theory with a positive $c_{cont}$. We call this case 'ferromagnetic' (in an obvious spin language) because the free energy is minimum for $\Phi_m = \Phi_n = \Phi$ ( =0 say; this is ferromagnetic because all 'spins' point in the same direction). This describes an s-wave superconductor. The continuum limit of the latter minimum is also well known. In the case of positive c(x,T), the mean field minimum is at $(\Phi_m - \Phi_n) = \pi$ so that we write $(\Phi_m - \Phi_n)$ as $\pi + (\delta_m - \delta_n)$ where $(\delta_m - \delta_n)$ is very small (natural in the continuum limit where m and n are again infinitesimally close to each other) and expand the latter to quadratic order in (m-n), we again have a Ginzburg Landau like third term. However, this is analogous to the 'antiferromagnetic' case in which the nearest neighbour moments tend to point opposite each other, and where such a Neel order is characterized by nonzero sublattice magnetization. In our case, such an arrangement of phases leads to a d-wave superconductor (the nearest neighbour (m,n) pair phases differ by 180°). Since this is what is observed, we take c(x,T) to be positive.

We note that the basic pairs are the same in both cases, and whether d-wave or s-wave superconductivity (extended s-wave in our case, because unlike strict s-wave superconductors, the onsite pair amplitude vanishes here due to strong on site repulsion or large positive Mott-Hubbard U) is favoured depends on the interaction between pairs. This is different from those approaches to superconductivity in the cuprates in which the form of the pairing potential is different for s and d wave symmetry superconductor. Here the same nearest neighbour pairs interact such that d-wave superconductivity emerges. This picture which seeks to differentiate pair formation and (long range) pair phase ordering is clearly most appropriate when the energy scales for the two are very different as happens for small doping x. For overdoped cuprates, this is not true, and a microscopic approach involving a pairing 'glue' characterized by an appropriate boson exchange function which leads to d-wave symmetry pairs may be more reasonable. Our theory is likely to be more natural in the underdoped, optimally doped and mildly overdoped regimes.

The pair interaction term can be thought of as a pair 'shift' term, namely that the pair at site m hops to the nearest neighbour site n. This can happen most clearly for strong correlations in the original square Cu lattice by the hopping of a hole from a site to the one diagonally opposite, i.e. to its next nearest neighbour in the original square lattice. This leads to a matrix element between $\psi_m$ and $\psi_n^*$. Indeed, a correlation between the size of this diagonal hopping term (generally called t') and $T_c$ at optimum doping has been noted by Pavarini et al[23]. Such a correlation, mysterious otherwise, is natural in the present approach, eg if microscopically there is a significant contribution to coupling between $\psi_m$ and $\psi_n$ (which coupling strength is proportional to $T_c$) via diagonal hole hopping (or t'). In recent work[24], a similar correlation between t' and 'fidelity', a quantity related to $T_c$, has been observed. We also point out that the functional we will use is not coarse grained but involves assuming that the nearest neighbour electrons form a spin singlet Cooper pair; this is supported by the fact that the d-wave pairing gap which emerges from our theory depends on **k** as {cos($k_x a$)-cos($k_y a$)} which is appropriate for nearest neighbour pairing and is in agreement with observations. The relatively small bare coherence lengths in cuprate superconductors ($\xi_o$ ~ 15A$^0$ and not ~10,000A$^0$ as in 'conventional' superconductors) also support this approach.

The Ginzburg Landau functional we propose for the cuprates is the following:

$$F = F(\{\psi_m\}) = \sum_m (a \Delta_m^2 + b \Delta_m^4) + c \sum_{mn} \Delta_m \Delta_n \cos(\Phi_m - \Phi_n) \qquad (2)$$

We assume for a, b and c the following:

$$a(x,T) = (f/T_o)^2 \{T - T_o(1-(x/x_c))\} \exp(T/T_o) \qquad (3\alpha)$$

$$b(x,T) = b_o (f/T_o)^4 T_o \qquad (3\beta)$$

$$c(x,T) = x\, c_o (f/T_o)^2 T_o \quad . \qquad (3\gamma)$$

The equations (3α) to (3γ) are characterized by one temperature scale $T_o$ which is specific to a cuprate as well as the dimensionless scale factors f and $b_o$, taken to be the same for all cuprates ($b_o$ might vary somewhat from material to material since it determines, e.g., the height of the specific heat hump that seems to be material dependent). This energy scale $T_o$ is approximately the temperature at which a(x,T) changes sign at x = 0. A particular cuprate has also a specific value of $c_o$. All the dimensionless quantities f, $b_o$ and $c_o$ are of order unity. We choose f=1.33 and $b_o$=0.1. For Bi2212, $c_o$=0.3 and $T_o$ (~ 400 K) reproduces the actual transition temperatures etc.. We are motivated by physical considerations and a desire to keep things as simple as possible in the above choice of a(x,T), b(x,T) and c(x,T); the choice is discussed below.

In Eq.3α, a(x,T) changes sign, for x = 0, at T=T$_o$ . This means that the thermodynamic probability distribution of a Cooper pair (ij) changes characteristically but smoothly around this temperature. We identify it with the 'bare' pseudogap crossover temperature T$^*_o$ which is our initial input estimate for T$^*$. In our theory, the average pair amplitude <Δ$_m$(x,T)> is not zero at any temperature because the probability distribution of Δ$_m$(x,T) is given by the Boltzmann factor exp(-βF) as mentioned above and is thus always nonzero. This probability distribution changes character as a(x,T) changes sign. There are preformed nearest neighbour bond spin singlet Cooper pairs (**not** preformed d-wave pairs) at all temperatures and their probability changes rather rapidly around T$^*_o$, but no long range order develops there. In the BCS mean field theory, T$^*$ is identified with the superconducting transition temperature. It so happens that because of large ξ$_o$ (or Cooper pair size) in conventional superconductors, fluctuation effects are small, the BCS mean field theory is quite accurate, and the true superconducting transition temperature (at which the system becomes stiff with respect to a change in the relative phase of ψ(**r**) at two points arbitrarily far apart) is very close to the BCS mean field temperature (i.e. in our language T* ≈ T$_c$).  In our case this is not so; the experimental T$^*$ ( if identified with the Cooper pair breaking crossover scale)  and T$_c$ differ dramatically especially for small x (T$^*$ >> T$_c$), while for x>x$_{opt}$ the distinction is not clear. Since both T$^*$ and T$_c$ vanish at x~x$_c$ ~0.3 we assume the simple linear extrapolation of (3α) where a(x,T) changes sign. This means that our input T$^*_o$ (x) has the straight line form T$^*_o$ (x)= T( 1 − (x/x$_c$)). We have also computed T$^*$ from our GL functional in several ways (see below). These results and the observed T$^*$(x) agree as to the size and the trend with x.

The term a(x,T) has a factor exp(T/T$_o$) exponential in temperature, which simulates the effect of the fact that fluctuations at temperatures well above T$^*$ ( but below the fermionic degeneracy temperature) are suppressed as is natural in any degenerate Fermi system. In its absence, the classical functional leads to the local gap <Δ$_m$(x,T)> = Δ$_{av}$(x,T)   becoming a constant at very high temperatures essentially because of the equipartition theorem applied to the classical degree of freedom ψ(x,T) with the above functional, as also shown by finite size simulations. In general, the quantity T$_p$ in the exponent  which scales T need not be the same as T$_o$; we assume T$_p$=T$_o$ for simplicity. Changes in it affect for example how rapidly Δ$_{av}$(x,T) increases as T decreases. The value chosen for b(x,T) is x independent and is (has to be) positive.

The proportionality of the coefficient c(x,T) to x is motivated as follows.  When Δ$_{av}$(x,T) is sizeable (as happens for T<<T$^*$), fluctuations in the Cooper pair magnitude  Δ$_m$(x,T) are negligibly small so that the phase stiffness temperature (determined basically by the third term in Eq.3) is of order z Δ$_{av}$(x,T)$^2$ c (where z is the number of nearest neighbours in plane). If we choose c α x, the well known Uemura correlation, namely that T$_c$(x) α x,  is reproduced . This is all selfconsistent and roughly true experimentally for small x < x$_{opt}$ in which regime T$_c$ << T$^*$. We assume, for simplicity, that c has this linear in x dependence for all x. In general, we could have taken a more complicated form for c(x,T), eg one going as x for small x, but differently for larger values. We however stick to this simple form which seems clearly indicated for small x. As mentioned above, it can be rationalized microscopically in strongly correlated systems as being due to the diagonal hopping of holes; the probability that this happens is proportional to the hole density x for small x. Assuming this simple form for x, the calculated $T_c$ is nearly linear in x for small x (x less than $x_{opt}$) and then decreases because thermal fluctuations in the magnitude $\Delta_m(x,T)$ are also significant in addition to the phase fluctuations, as well as because of the fact that fluctuations in $\Delta_m(x,T)$ and in the phase $\Phi_m(x,T)$ are strongly coupled via the c term. We obtain as a consequence the well known parabolic curve for $T_c$ (Section III and ref.2).

The postulated Ginzburg Landau form Eq. 2 with a,b and c as described in Eq.3 is the basis on which we calculate a large number of physical properties of the cuprates analytically in various approximations eg single site mean field theory and cluster mean field theory, as also numerically for example via finite size Monte Carlo simulations (Sections III and IV). We also use where indicated, the 2d Berezinskii Kosterlitz Thouless (BKT) theory of thermal vortex unbinding. We compare our results with experiment as shown below (see also Refs. 2 and 3).

Literally, such a form for F(ψ) with only low order powers in ψ is expected to be accurate only when $\Delta_{av}(x,T)$ is small (namely around $T^*_o$). However, because it has the right form for small ψ as also qualitatively the right form for large |ψ| (i.e. it increases for large |ψ| and is thus a stable probability distribution) we use it over a much wider regime of x and T (see eg Fig.4). This is somewhat like the case of the Ginzburg Landau theory for 'conventional' superconductors which is expected to be accurate near $T_c$ (~ $T^*_o$ in that case) but has been subsequently used over a much wider range of temperatures. Also, in our focus on the Cooper pair degrees of freedom $\psi_m(x,T)$ and on superconductivity, we have not made explicit other (bosonic as well as fermionic) degrees of freedom which might be significant for different (x,T), and might be system specific, and might be coupled with $\psi_m(x,T)$ (see Section V).

## III. Some results using the new Ginzburg Landau functional

### a) Local gap and the pseudogap temperature

Values for the thermodynamic average $\langle\Delta_m(x,T)\rangle = \Delta_{av}(x,T)$ namely the average local gap magnitude calculated using the functional of Eq.2 are shown in Fig.5. These have also been obtained numerically using equations 2 and 3, for finite sized square lattice systems. As expected from the probability distribution which is proportional to exp(-βF) where F is given by

Eq.2, $\Delta_{av}(x,T)$ is nonzero at all temperatures, but rises as T decreases. Its maximum slope $T_{ms}^*$ is operationally identified with $T^*$ and is plotted in Fig. 6 as a function of x. It is seen to decrease as x increases. (The rapidity of the rise in $\Delta_{av}(x,T)$ as T decreases depends on $T_p$, the energy scale factor of the exponential term in a(x,T); this is shown in Ref.2). Because of the abrupt change in $\Delta_{av}(x,T)$ at $T_c$, $\Delta_{av}(x,T)$ has an infinite slope there and thus there is always a second $T^*_m$ close to $T_c$. We follow $T^*_m$ from x=0 as it decreases with x, 'touches' $T_c$ at $x=x_{ps}$ and use the latter branch beyond it. This essentially goes along the $T_c$ curve(Fig.6). On the other hand, if the c term which leads to long range phase coherence is neglected and only the first two terms in F (from Eq.2) are used to calculate $\Delta_{av}(x,T)$, $T_m^*$ decreases for x beyond $x_{ps}$, 'goes through' the superconducting dome, touching T=0 at $x=x_{ps}^0$ and continuing to be very nearly at T=0 beyond this x (Fig.6). We believe that this finding is behind the observation of differently behaved pseudogaps[23], namely the former in one type of measurements and the latter in those which destroy superconductive phase coherence (eg via perhaps a magnetic field, or via a small concentration of impurities such as Zn). From our results it appears that if $x_{ps}^0$ is identified with a quantum critical point (or QCP, which, though hidden from experiment, is widely believed to exist and to have significant effect on observables), it is one in which the phase degrees of freedom are somehow suppressed.

b) **Superconducting transition ( $T_c$)**

The superconducting transition sets in when the long wavelength phase stiffness or the superfluid density $\rho_s$ becomes nonzero. One circumstance under which this happens is broken symmetry, namely $<\psi_m> \neq 0$ . Thus location of broken symmetry is one way of finding $T_c$ (which we do in d=2 as well, using Eq.2, though strictly, because of logarithmically divergent thermal order parameter fluctuations, such a broken symmetry is not possible; this is the famous Mermin Wagner theorem. However, our mean field estimates are indifferent to this). We have used single site as well as cluster mean field theory to find $T_c^{MF}$ and $T_c^{CMF}$ respectively (Fig.7). For d=2 as in our planar case, we have numerically calculated $\rho_s$ in the standard manner mentioned below, and the Kosterlitz Nelson condition[25] $\{\rho_s(T_{BKT})/T_{BKT}\} = \{2/\pi\}$ to locate $T_{BKT}$ or the vortex binding temperature (Fig.7). The latter is known[25] to be below the former typically by a factor of ~1.5 .

The phase stiffness or superfluid density $\rho_s$ is the second derivative of the free energy with respect to the vector potential **A** in the limit of zero **A**. The vector potential enters the functional Eq.2 via the usual Peierls factor which is added to the phase, in the third term of Eq.(2). The general expression for the superfluid density is thus

$\rho_s = - (c/2\, N_b)<\sum_{mn}\Delta_m\Delta_n\cos(\Phi_m-\Phi_n)> - (c^2/2N_bT)<(\sum_{mn}\Delta_m\Delta_n\sin(\Phi_m-\Phi_n))^2>$   (4)

where $N_b$ is the number of sites in the bond centre lattice ( $N_b=2N$ where N is the number of sites in the original lattice). We have used finite system Monte Carlo simulations and the definition of $\rho_s$ as in Eq.4 to estimate it. The effect of the neglected weak interlayer coupling is expected to be small so that our results can be meaningfully compared with experiment as necessary. The parabolic shape we obtain is universally observed in the cuprates. We can even get quantitative agreement by fixing the parameter $c_o$ of the functional (see the discussion after Eq.3).

A widely used dimensionless parameter for superconductors is the ratio $(2\Delta_o/k_BT_c)$ where $\Delta_o$ is the single particle energy gap at T=0 (In the d-wave case, $\Delta_o$ is the antinodal gap). This has the value 4.3 for a d-wave superconductor in BCS mean field theory. Not surprisingly, our choice of parameters a, b and c leads to this ratio if we use the BCS mean field estimate $T_o^*$ in the above equation instead of $T_c$. Our calculations give a value for it in the range 9-10 over a wide range of doping around $x_{opt}$, in agreement with experiment and in contradiction with the widely fulfilled expectation in conventional superconductors, for which 3.76 is the well known appropriate value for an s-wave superconductor.

The parabolic shape we find for the $T_c$ curve is expected to be unreliable near both ends[26], where $T_c \rightarrow 0$, so that quantum fluctuations are particularly important. Here our characterization of the Cooper pair degrees of freedom in terms of a classical (ie time independent) $\psi_m$ is specially inadequate. At other values of x, and at nonzero temperatures, the classical approximation works well. We use in Section V a simple quantum generalization of Eq.2 which leads to a $T_c(x)$ curve which is in quantitative agreement with experiment near both ends of the $T_c$ curve for say $La_{2-x}Sr_xCuO_4$.

c) **Superfluid density $\rho_s(x,T)$**

We have used Eq.4 to obtain $\rho_s(x,T)$ numerically. Some features are mentioned in Ref.2. One of them is the following. The superfluid density or phase stiffness $\rho_s(x,0)$ at zero temperature is found to have a 'boomerang' shape as a function of the superconducting transition temperature $T_c$ we calculate, eg in cluster mean field theory. Namely, it first increases as $T_c$ increases (Uemura correlation), has a maximum, and falls as $T_c$ decreases again. This behaviour has been widely observed experimentally[27]. In BCS theory, the superfluid density is very large; in pure systems it is proportional to electron density while $T_c$ is exponentially small so that such a shape is not expected. The fact that our calculation leads to a pattern for the zero temperature superfluid stiffnes as a function of $T_c$ which is close to that actually observe, while the BCS behaviour is qualitatively different, is another indication of the importance of phase fluctuations in determining both $T_c$ and $\rho_s(x,T)$.

d) **Specific heat**

Equation 2 describes a classical statistical mechanical system with $\psi_m(x,T)$ as the only degree of freedom. We have calculated the specific heat of such a system; it is due to thermal Cooper pair fluctuations. The cuprate actually has a large lattice specific heat, and also electronic specific heat arising from unbound Cooper pair electrons. On subtracting these (see Ref.2 and Ref.4 for details), we find a sharp peak in $C_v(x,T)$ at $T_c$, and also a broad bump centered around the pseudogap temperature. The separation is most clear for underdoped cuprates. The peak is associable primarily with phase fluctuations which are involved in the transition to superconductivity, almost exclusively so for underdoping or small x. The bump at higher temperatures is connected with the order parameter magnitude degrees of freedom $\Delta_m(x,T)$ which are thermally excited around these temperatures. Since there is no phase transition at $T^*$, this feature is not a sharp peak but a rounded bump. We compare our results with both direct observations of the specific heat[29] and indirect ones[30] (these use accurately measured thermal expansion and a Maxwell relation connecting it with the specific heat). The former, namely the direct ones, done over a wide range and after subtraction of lattice and other electronic specific heat, have exactly the same features. For example, the specific heat peak height at $T_c$ calculated by us is compared with that measured for YBCO[29] as a function of doping, and we find good agreement. We have also calculated the effect of a magnetic field on the specific heat. This is done by introducing a Peierls phase factor term in the argument of the cosine in the third term of Eq.2. The size of the term is proportional to the magnetic flux in a plaquette and thus directly to the applied magnetic field. We observe a rounding of the specific heat peak (Fig.8), anomalous in that a very small flux (in units of the flux quantum in a plaquette) produces significant rounding. Our results are again in good agreement with recent high resolution measurements[30] of specific heat in a magnetic field.

e)    **Vortex**

We have obtained the structure of a vortex at T=0 from our theory. At T=0, Eq.2 describes the energy of the cuprate as a function of $\psi_m(x, T=0)$. We introduce a single vortex into the system by requiring that around a particular square plaquette the phase Φ winds around by 2π. If this phase change is equally distributed on the four corners of the plaquette, we say that the vortex core is at its centre, equidistant from them. On any closed curve which passes through the lattice points (the only physically meaningful points here) and surrounds the core, the phase changes by 2π on one circumambulation. With this constraint, and with the constraint that the magnitude of the order parameter attains its uniform system value asymptotically far from the core, we numerically minimize the energy as a function of $\Delta_m$ and $\Phi_m$ at each bond centre lattice site. The results are plotted in Figure 9. We see that the nature of the vortex changes as x increases. In the underdoped region (small x, eg x=0.1) the order parameter magnitude does not decrease greatly as one approaches the core radially; one has essentially a phase or Josephson vortex, something possible only in a lattice (or granular) system. For large x (overdoped, eg x=0.3), the order parameter magnitude decreases considerably as one similarly approaches the core; the topological defect is more like a BCS vortex, for which the order parameter magnitude vanishes at the core.

## IV. Electron spectral function and ARPES

We have described above a phenomenological approach to Cooper pair fluctuations in the cuprates and its many experimental consequences. These seem to be in good agreement with observations. However, a large number of experiments involve the coupling of these degrees of freedom to electrons, photons, neutrons etc.. For example, in Angle Resolved Photo Emission Spectroscopy (ARPES), a technique that has seen revolutionary advances largely due to the impetus provided by the cuprates, one measures the energy spectrum of electrons with a definite planar momentum kicked out of the solid by an incident photon (this is a highly developed version of the photoelectric effect). In the last decade or so, high resolution ARPES experiments on cuprates have unearthed many interesting features. Early experiments[11] showed that the superconducting gap magnitude is a strong function of **k** on the Fermi surface, and is of the form $|\cos(k_x a) - \cos(k_y a)|$ where $k_x$ and $k_y$ are the x and y components of the in plane momentum on the Fermi surface. Since then, strange features like Fermi arcs[10] above $T_c$, and the 'bending' of the gap below $T_c$[14] have been discovered.

We show here that these features (and others) follow from the inevitable quantum mechanical coupling of low excitation energy (near Fermi energy) electrons to long wavelength Cooper pair fluctuations which are formed of such electrons. Thus as the d-wave like symmetry correlation length $\xi$ of the pairs diverges for $T \rightarrow T_c$, electrons move in an annealed homogeneous medium of superconducting patches of size ~ $\xi$, disappearing into them and reappearing. The patches themselves are transient. This is different from the picture of preformed d-wave symmetry pairs with a lifetime. We see below that it leads to a characteristic, inevitable self energy for the electrons and unusual electronic phenomena observed in ARPES. In order to access these, one needs to have a theory of electronic excitations in cuprates. At least in strong correlation electron theory, this **is** the basic electronic problem for which there is no paradigm, and presumably from which the behaviour of the Cooper pair and other degrees of freedom and their mutual coupling, also follows. Lacking this, we have proposed and developed above a phenomenological description of the free energy as a function of only pair degrees of freedom. We propose now a halfway or semiphenomenological solution which explicitly has both the Cooper pair degrees of freedom, and the low energy electron degrees of freedom. We assume that there are low energy quasiparticles in the cuprates . There is strong evidence in favour of this assumption from simulations of the 2d Hubbard model[31]. The effective low energy one electron Hamiltonian we assume is

$$H = -\sum_{i,j,\sigma} t^*_{ij}\, a^+_{i\sigma}a_{j\sigma} - J^*\sum_{<mn>}\, [\psi_{mn}\, a^+_{i\downarrow}a^+_{j\uparrow} + h.c.] \qquad (5)$$

where $t^*_{ij}$ is the renormalized electron hopping amplitude between lattice sites i and j for an electron with spin index σ. . The second term describes a pair moving in the complex potential $\psi_{mn}$, with <mn> denoting the nearest neighbour sites in the bond lattice. The relation between the pair mn and the pair ij (both nearest neighbours) is shown in Fig.4. $J^*$ is the effective, AF exchange determined, pair attraction. The dynamics of $\psi_{ij}$ is determined by the functional Eq.2 (though this particular assumption is not necessary for most of the results described below).

The quasiparticle propagation is affected by Cooper pair fluctuations as follows. The quasiparticles hop from site to site on the square Copper lattice, become a Cooper pair and unbecome one, so that there is a quantum mechanical admixture between an electron, a Cooper pair, and another electron of nearly the opposite momentum. This leads to a self energy $\Sigma(\mathbf{k},iv_n)$ for an electron of momentum $\mathbf{k}$. The Feynman Matsubara diagram for it is shown in Fig. 9. It is the well known bosonic fluctuation exchange process, in which the fermion exchanges a bosonic fluctuation (in this case a Cooper pair) . (It is standard for say coupled electron phonon systems). We have calculated this self energy and the consequent electron spectral density $A(\mathbf{k},\omega) = -(2/\pi)$ Im $[G(k,iv_n \rightarrow \omega + i\delta)]$ using the standard Dyson equation for G, namely $G^{-1}(\mathbf{k},iv_n) = G_o^{-1}(\mathbf{k},iv_n) - \Sigma(\mathbf{k},iv_n)$. The effects of strong correlation on the quasiparticle can be approximately included via the Gutzwiller projection approximation[32]. At low temperatures, this factor reduces the bare one electron hopping $t_{ij}$ by a multiplicatory term $(2x/(1+x))$ ( so that $t^*_{ij}= (2x/(1+x))\, t_{ij}$ ) which accounts for the prevention of double occupancy of a site (for U=∞). The electron propagator using Eq.5 has the correlation renormalized bandwidth, but the quasiparticle residue $z_k$ = 1. We can include the effect of interactions which reduce $z_k$ from this value by using a propagator $z_k(iv – t^*_k)^{-1}$ for the quasiparticle in the intermediate state (Fig.10). For a weakly $\mathbf{k}$ dependent $z_k$ (as is seen to be the case from simulations[31]) this factor can be absorbed in our calculations as renormalization of an unobserved quantity. The self energy Fig.9 also involves the Cooper pair propagator. For small momenta near $T_c$ this propagator has a universal form. Thus, this self energy is inevitable. It is also quite generally calculable.

It turns out that the leading vertex correction to the self energy diagram is identically zero above $T_c$, and involves the Goldstone mode (or spin wave) connected with the LRO below $T_c$. This latter is of order $(T_c/T_F)\sim 10^{-2}<<1$ where $T_F$ is a characteristic electron degeneracy temperature. Thus a Migdal like theorem obtains here, and vertex corrections can be ( and have been) ignored. In our calculations, we have used the bare electron propagator in the intermediate state in Fig.10, rather than the self consistent or true propagator. While we can calculate quantities using the latter, experience with electron phonon systems shows that it is adequate to use the bare one. For the Cooper pair propagator there, we use a general real space form, which for large interpair distances $|\mathbf{R}| = R$ goes like $R^{-n}\exp(-R/\xi)$ after the LRO term is removed. Here $\xi$ is the (d-wave symmetry) pair correlation length. We find that the self energy due to coupling with long wavelength Cooper pair fluctuations leads to a non Fermi liquid, which is moreover particle hole asymmetric. There is ARPES evidence for both.

We use the BKT form[25] for the temperature dependence of $\xi$ eg in the underdoped regime where the functional of Eq.2 reduces at temperatures of interest to one of essentially fixed length 'spin' with direction $\Phi_m$ as degrees of freedom at temperatures of interest. A synoptic representation of the spectral density for different values of doping and of temperature is shown in Fig.11. We see that below $T_c$, the single peaked structure (with the peak at the Fermi energy) is found only at the node. For all other angles, there are two peaks; the separation between them is twice the gap. We have plotted this gap as a function of position on the Fermi surface, this latter being labeled by $|\cos k_x a - \cos k_y a|$ ( Fig. 11) and notice that there is a 'bending' with respect to the canonical d-wave form $\Delta_k = \Delta_o |\cos k_x a - \cos k_y a|$ which becomes more pronounced as T rises towards $T_c$. This is due to the self energy term arising from the coupling of the fermion with the thermal collective excitations of the order parameter fluctuations (Fig.10). In the literature[14], the bending has been associated with two gaps, one antinodal and one nodal. We see that a model with only one gap, whose LRO or mean field form is of d-wave symmetry, namely $(\cos k_x a - \cos k_y a)$, and with quasiparticles necessarily coupled to pair fluctuations explains the observations qualitatively and quantitatively. One of our other results is that below $T_c$, the antinodal spectral function does develop a coherence or quasiparticle peak. We have calculated its strength, and find that it increases with superfluid density or phase stiffness (as is observed).

We also see from Fig.11 that above $T_c$ there is a region in **k** space on the Fermi surface, for which the spectral density has only one peak, at the Fermi energy. This region increases in extent as T increases; it is the Fermi arc. A visually compelling picture of this is shown in Fig.12. From this point of view, $T^*$ is defined as the temperature at which the arc expands to the zone edge (this T* is also shown in Fig.6). We have also plotted the reduced arc length as a function of reduced temperature, and find that it agrees with the nearly linear dependence observed (Fig.13). A simple criterion, inspired by our self energy form, and one which is physically reasonable, is to identify the arc edge with the **k** point on the Fermi surface which satisfies the condition $\Delta_k = v_k/\sqrt{2}\xi$.

We have also calculated a number of other features implicit in the spectral density results. For example, the pseudogap is characterized by there being both a two peaked structure in it with a certain peak density, **and** a nonzero density at the Fermi energy. This property is quantified by a function L , namely the loss of low energy spectral weight L(Φ) , which is the ratio of the latter to the former. L(Φ) is zero for a real gap, and unity for no gap. At different temperatures (for a given doping) we find intermediate values of L(Φ) which are in agreement with experiment.

## VI. Discussion and Prospect

We have proposed and developed above a phenomenological theory of phenomena in cuprates in a wide temperature and hole doping regime where high temperature superconductivity occurs. We describe the free energy of the system as a functional of the complex quantity $\psi_{ij}$ = $\psi_m$ = $\Delta_m \exp(i\Phi_m)$ where i and j are nearest neighbour sites on the square Cu lattice. The functional is Ginzburg Landau like, namely it is identical in form with that used by them in their phenomenological theory of 'conventional' superconductivity, though it is quite different in content. It leads to d- wave symmetry superconductivity as is observed in cuprates in contrast to the s-wave superconductivity of conventional superconductors, to a regime of uncondensed nearest neighbour pairs, to a pseudogap crossover temperature $T^*$ which has the observed hole doping or x dependence and to the observed parabolic $T_c$ (x) curve. Our approach can be characterized by saying that in it, both the Cooper pair magnitude $\Delta_m$ and the phase $\Phi_m$ are important coupled low energy degrees of freedom. The coupled thermal probability distribution for them is given by exp{-βF} where F = $\sum_m$ (a $\Delta_m^2$ + b $\Delta_m^4$) + c $\sum_{mn} \Delta_m \Delta_n \cos(\Phi_m - \Phi_n)$ (with 0<$\Delta_m$<∞ and 0<$\Phi_m$<2π, these being the ranges of $\Delta_m$ and $\Phi_m$ respectively ). Especially for small x, and with our choice of a, b and c, the probability distribution for $\Delta_m$ is dominated by the first two terms in F, leading to a crossover or pseudogap temperature below which <$\Delta_m$> is sizeable. The superconductivity transition is a long range order transition in the phase brought about by the short range interaction between $\psi_m$ and $\psi_n$ described by the third (phase interaction) term in F. For larger x, the coupling between $\Delta_m$ and $\Phi_m$ is important so that $T_c$ is no longer determined mainly by phase fluctuations $\Phi_m$.

The fact that (Cooper pair) phase fluctuations are important in cuprate superconductivity has been brought out most clearly in the extended experimental work of Ong and collaborators on the Nernst effect[33] and fluctuation diamagnetism[34]. A few years after the discovery of high temperature superconductivity, Emery and Kivelson[35] pointed out through an analysis of the small superfluid stiffness $\rho_s$ in the cuprates and comparison with the large $\rho_s$ in

'conventional' superconductors that phase fluctuations are crucial for their superconductivity. This is specially clear for small doping. As hole doping increases, we see from our phenomenological theory that the free energy cost of pair magnitude fluctuations comes down, and that the coupling between these and the phase fluctuations determines $T_c$. This regime is closer to the BCS case, which can be thought of as the limit in which the (coupled) nonzero magnitude transition and the phase stiffness transition occur at the same temperature.

In the spin language, we associate with $\psi_m$ a spin $\mathbf{S}_m$ of length $|\mathbf{S}_m|$ (or pair magnitude $\Delta_m$) pointing at an angle $\Phi_m$ with the x axis ($\Phi_m$ is the pair phase). The free energy F of Eq.2 can then be regarded as the sum of two terms namely $F = F_o + F_1$. The first one depends only the local spin length and generates its thermal probability distribution , ie $F_o(\mathbf{S}_m) = \sum_m (a\, \mathbf{S}_m^2 + b\, \mathbf{S}_m^4)$ and the second is a spin interaction term $F_1\{\mathbf{S}_m\} = c \sum_{m,n} \mathbf{S}_m \cdot \mathbf{S}_n$. For cuprates, the 'spin' formation (crossover) temperature is $T^*$ and is largely determined by $F_o$ in the underdoped regime. The long range order temperature (determined in the same doping range by the interaction term $F_1$) is $T_c$. In our case, the interaction is such as to lead to Neel or antiferromagnetic order (which translates into d-wave superconductivity). The two temperature scales seem well separated only for underdoped hole cuprates. Beyond optimum doping they are not, and the degrees of freedom (length and angle or magnitude and phase) are strongly coupled. When the spin formation temperature and spin ordering cannot be separated, one can have the BCS limit which is natural for itinerant uncorrelated fermions. The cuprates are unusual; in an itinerant fermionic system, one goes, as the hole doping increases, (in this point of view) from the limit of ordering of well developed spins to the limit of long range spin ordering without a well defined spin formation temperature above it. In most (insulating) spin systems however, the spins are formed already at far higher (crossover) temperatures ($T_s^f$) than those at which their interaction leads to long range order at $T_s^{LRO}$, so that the analogue of the pseudogap temperature $T^*$, namely $T_s^f$ is much larger than the analogue of the superconducting $T_c$, namely the spin ordering temperature $T_s^{LRO}$.

The relevant low energy degree of freedom $\psi_m$ in our theory above is classical, namely it has no dependence on 'time' $0<\tau<\beta$. This is a good approximation at not too low temperatures, which are generally of interest to us in our calculations. It is qualitatively incorrect at both the underdoped and the overdoped ends of the $T_c(x)$ curve, and the time dependence of $\psi_m$ is crucial. We have investigated the effect of quantum phase fluctuations on $T_c$ by adding to F in Eq.2 a nearest neighbour charge interaction term $(1/2) \sum_{mn} q_m V_{mn} q_n$ where the charge $q_m$ and the phase $\Phi_m$ are well known to be canonically conjugate, so that this term causes the phase to be dynamic. We take $V_{mn} = V_o\, \delta_{mn}$ if m and n are nearest neighbours, and to be zero otherwise. In our calculation, we have ignored the long range of the Coulomb (or charge) interaction, as well as ohmic dissipation. It has been argued[36] that these two factors together result in a fluctuation spectrum similar to that obtained in an approximation which ignores both, and describes the effect as that of a short range interaction (nearest neighbour in our case). With $V_o$ = 0.09$T_o$, we have calculated the $T_c$ curve for LSCO and have compared it with experiment in Fig. 14. Quantitative agreement is obtained with a choice of parameters almost identical with that mentioned in Section II. Noteworthy are the facts that $T_c$ vanishes at about x=0.05 at the lower end (not at x=0 as is the case for classical phase fluctuations ), and at x~0.28 (not at x~0.35) at the upper end. The general parabolic shape is more symmetrical.

The orbital effect of a magnetic field on superconductivity enters through the Peierls phase factor ∫**A**.d**l**, involving the vector potential **A** and added to the phase term say in Eq.2. We have not explored its effects (except showing that it will round the specific heat curve for surprisingly small magnetic fields, Section III). In particular, our Ginzburg Landau theory is expected to lead to a reduction in $T_c$ because of this term, and to a new interpretation of the bare coherence length $\xi_o$ (not necessarily the Cooper pair size) in terms of the physically reasonable and dimensionally correct equation $(1/T_c)(dT_c/dH) = (\xi_o^2/\Phi_o)$ where $\Phi_o$ is the flux quantum.

We have ignored other bosonic and fermionic degrees of freedom which may be relevant for the low energies and temperatures of interest to us. They may couple to the pair degrees of freedom which are the only ones considered here. Perhaps the most important is the spin degree of freedom, basically bosonic; in the undoped cuprate, antiferromagnetic LRO implies that the Cu spins are strongly coupled, while on doping with holes, one very quickly ( for x~ 0.04) has mobile holes and superconductivity, and the AF correlation length at x~0.1 is small, ~three lattice constants. How this crossover happens is one basic question we have not investigated. There are also phenomena such as stripes[37], electronic nematicity[38], checkerboard order[39] and possible T-reversal invariance breaking[40] which (by themselves and because of the coupling of these to Cooper pair degrees of freedom) can lend themselves naturally to a Ginzburg Landau description. We have further assumed that the fermionic (or electronic) degrees of freedom, when integrated out in favour of electron (Cooper) pair or bosonic degrees of freedom give a simple functional Eq.2., even though unpaired electronic excitations are also a low energy degree of freedom and so should be kept explicit. One thus has a two interacting quantum fluid system, one bosonic (treated here as a classical fluid) and another fermionic. We have taken into account the separate effect of fermionic degrees of freedom where necessary, eg in our calculation of the specific heat (Section III) where their contribution is subtracted, or in their coupling with the bosonic pair fluctuations (Section IV). A phenomenological theory such as that above raises the obvious, and perhaps the most significant question in the field: what is the appropriate microscopic electron theory in the presence of strong repulsive local correlations and how does one go from a collection of interacting, fluctuating, spins to a fluctuating Cooper pair system of the above sort with increasing hole doping, in the background of such strong correlations? One aspect of this question, of great current activity (and also not addressed here), is the observation of periodic quantum oscillations[41] in equilibrium and transport properties of very clean cuprates at very low temperatures in the superconducting phase, as a function of a large magnetic field perpendicular to the planes. We have also not considered the effects of static disorder (eg in the pair phase or one electron charge potential), and of lattice displacements (coupled to electrons).

In summary, we believe that high temperature superconductivity in the cuprates provides us with a window through which we glimpse the strange quantum world of strongly correlated electrons (constituting a strongly interacting Fermi field on a lattice), and organized in a certain hole doping regime and some systems into interacting Fermi (single particle) and Bose (Cooper pair) degrees of freedom. We have described above a theory of the goings on related to superconductivity, namely the latter . The system is phenomenologically described in terms of an interacting classical Cooper pair field $\psi_m$. We find among other things that the observed d-wave symmetry superconductivity emerges as result of interaction involving the $\psi_m$s , and are able to make sense of a wide range of puzzling observations.

## Acknowledgments:

I am very thankful to Professor Phua and other organizers of the conference on 'Emergence in Field Theory' (August 4-6, the Institute of Advanced Study, Nanyang Technological University, Singapore) for giving me an opportunity to summarize our work on this new departure. I am grateful to my collaborators Dr Sumilan Banerjee and Professor Chandan Dasgupta for our journey together. I am thankful to supporting organizations, namely the Department of Science and Technology, New Delhi, India, and the National Centre for Biological Sciences, Bangalore, India. This article could not have been written without the help of Dr Sumilan Banerjee. My thanks to him.

Figures.

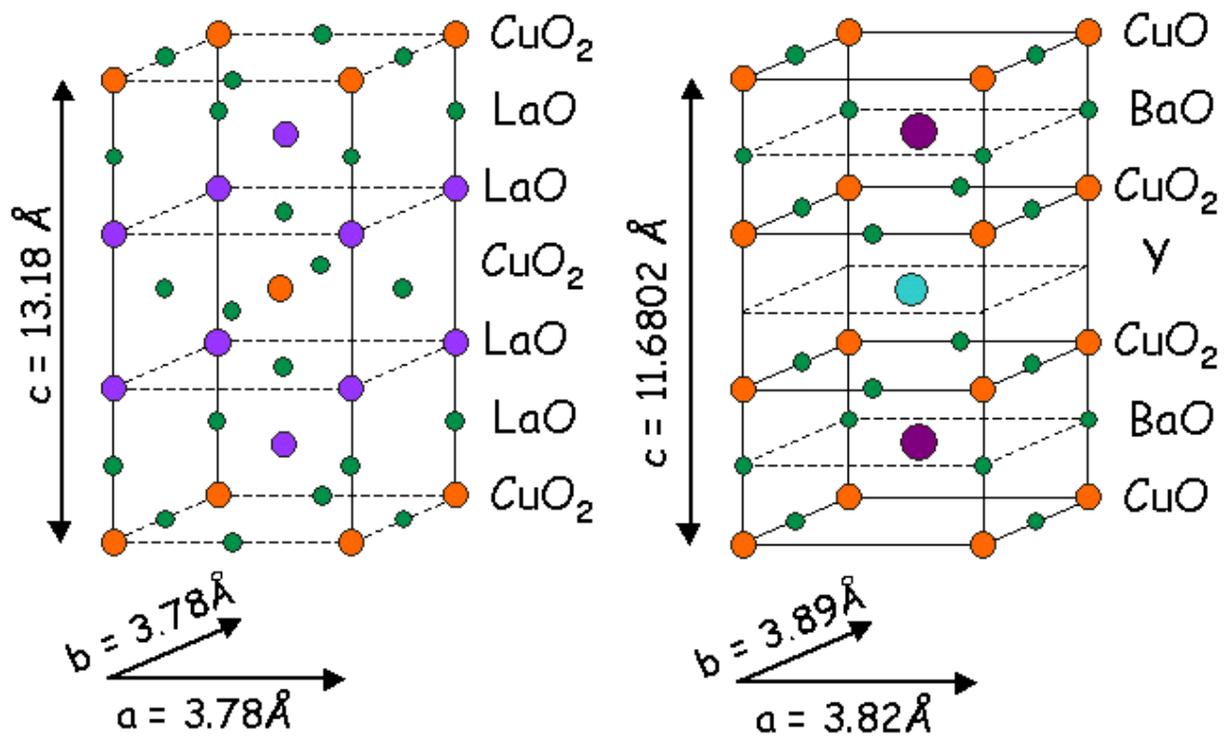

FIG. 1: Crystal structures of LSCO and YBCO.

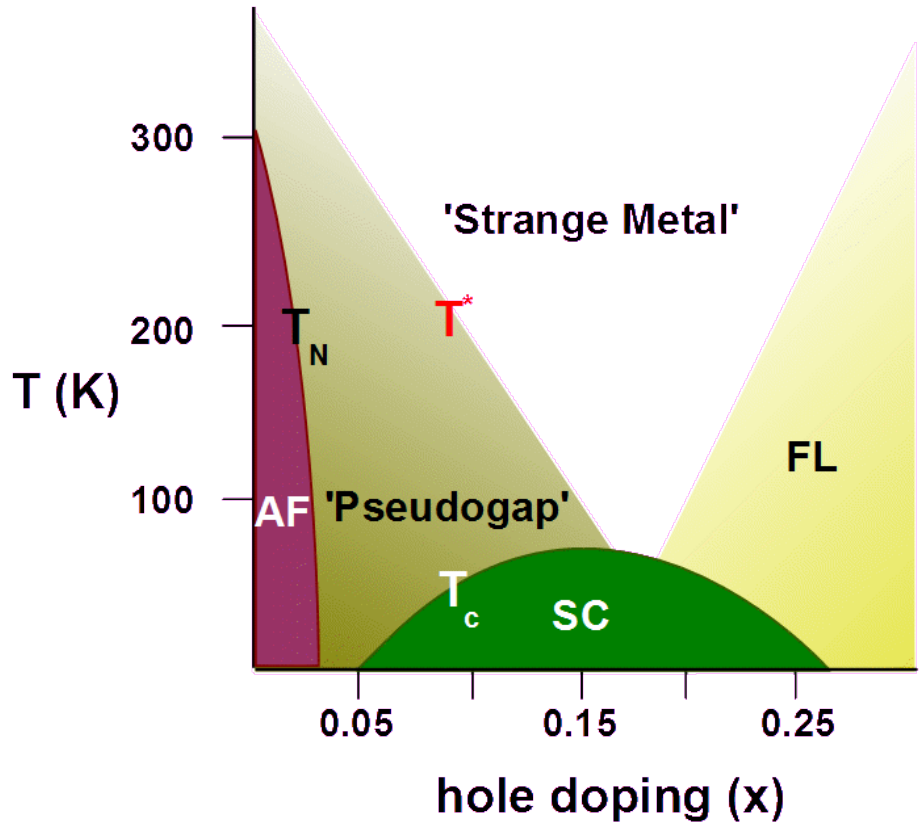

FIG. 2: Universal 'phase diagram' of hole doped cuprates in the hole doping, temperature or (x,T) plane.

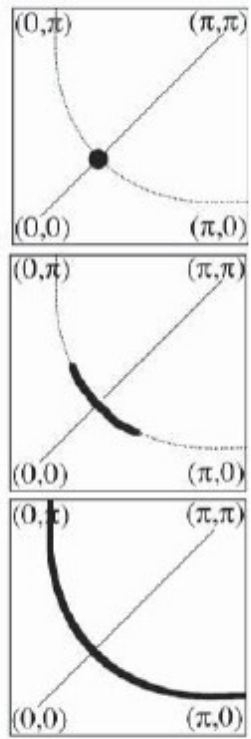

FIG. 3: Fermi surface plus Fermi arc (from the review by J. C. Campuzano, M. R. Norman and M. Randeria in The Physics of Superconductors, vols. I and II , edited by K. H. Benneman and J. B. Ketterson, Springer, New York, 2003). The three schematic figures denote, respectively, the Fermi arc at T=0 (shrunk to the nodal point), the arc at a temperature above Tc but below T*, and the same at T*.

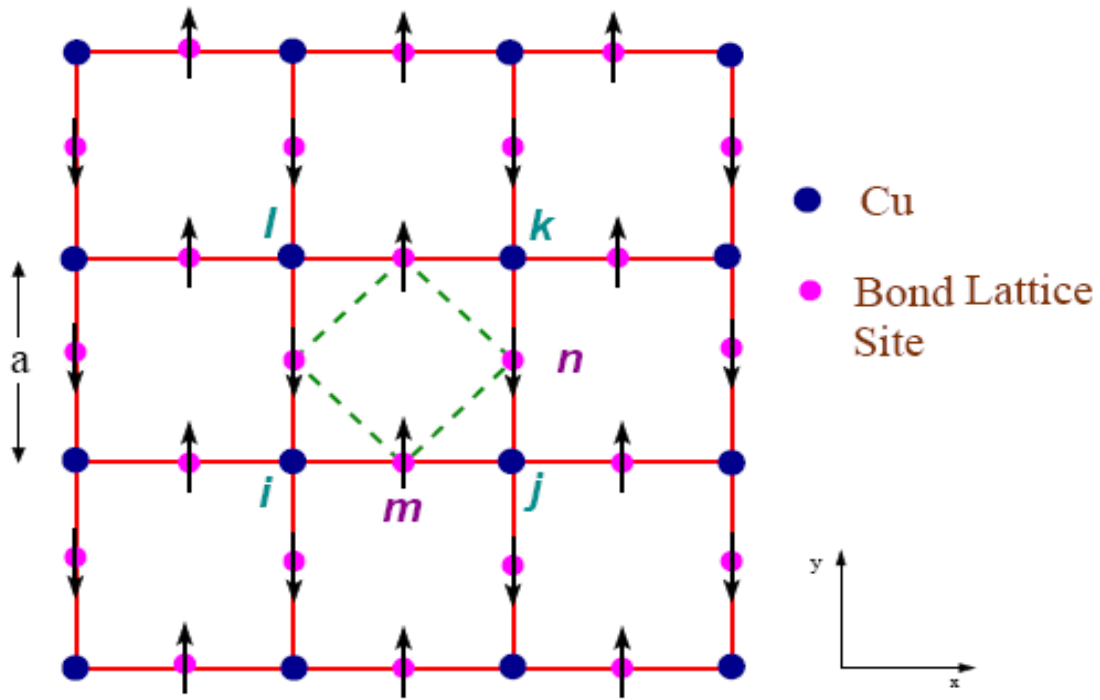

FIG. 4: The basic square planar lattice of the cuprates, with the location of Cu atoms in blue and that of O atoms in magenta. The arrow represents the direction of the equivalent 2d-XY spin.

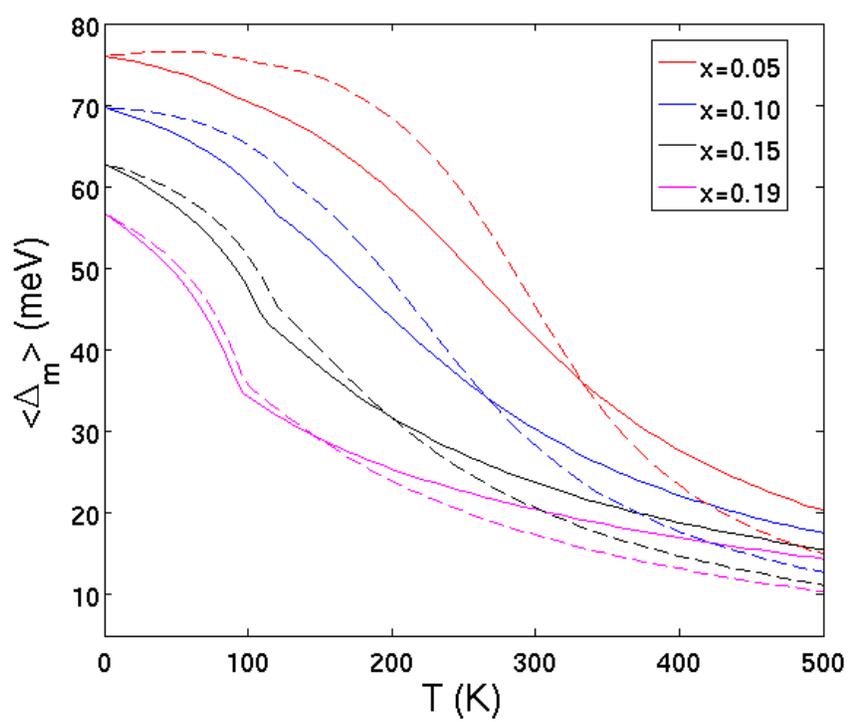

FIG. 5: Average local gap as a function of T for different values of doping x.

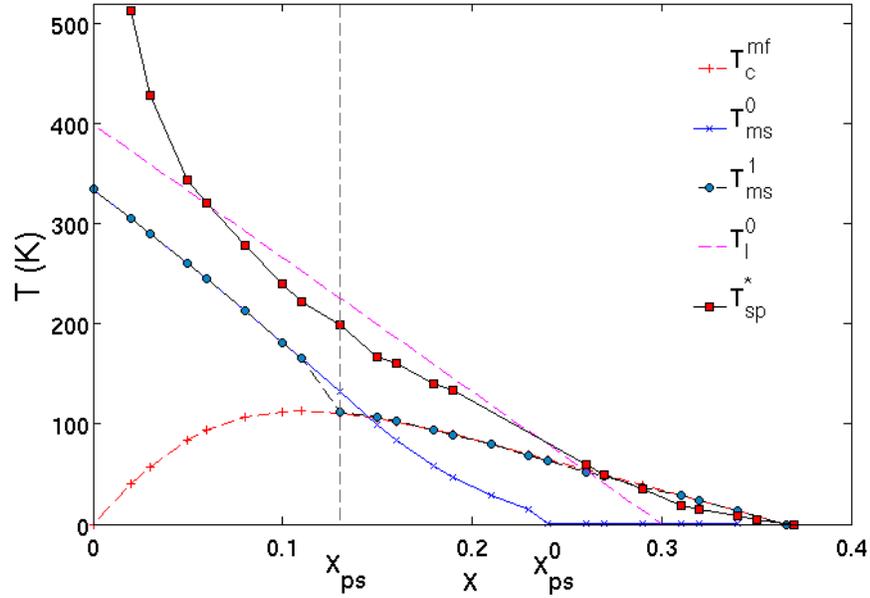

FIG. 6: Various T* s namely T*$_o$ ( shown as $T^{0}_{l}$) , the two pseudogaps  $T^{0}_{ms}$ and $T^{1}_{ms}$ from the maximum slope criterion, and T*$_{sp}$ from nodal gap filling. The increase in the last for small x is an artifact due to the neglect of spin fluctuation degrees of freedom etc.. The $T_c(x)$ curve is shown to set the scale.

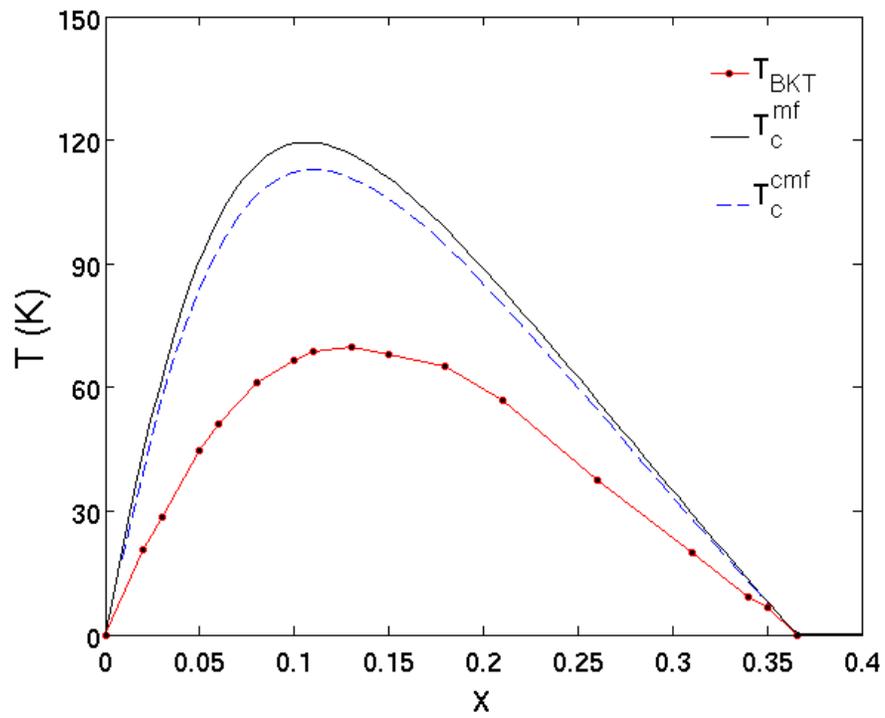

FIG. 7: $T_c(x)$ in single site mean field theory $T_c^{mf}$, cluster mean field theory $T_c^{cmf}$, and $T_{BKT}$.

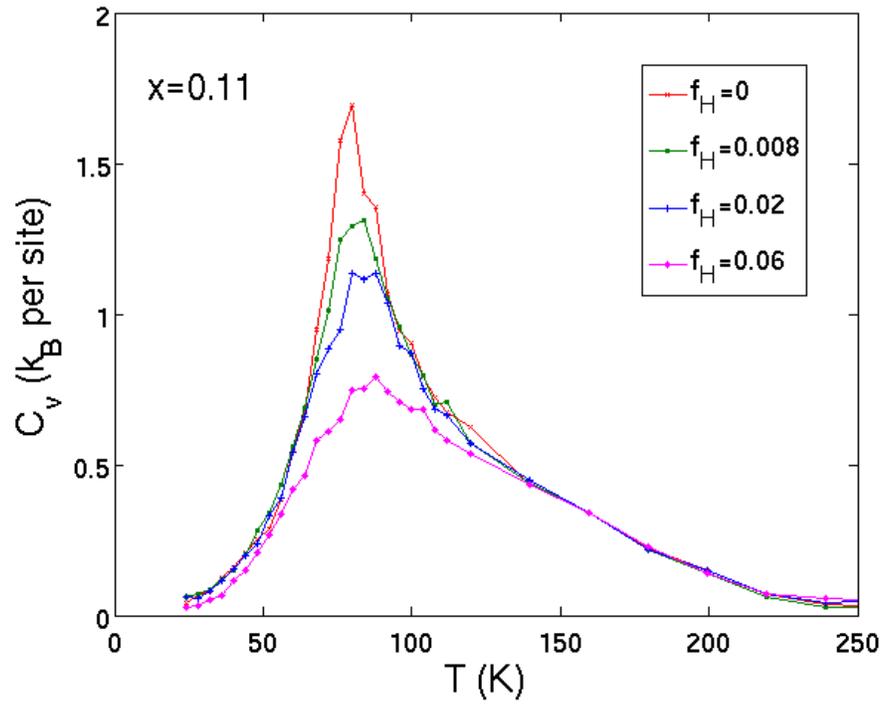

FIG. 8: Specific heat peak with and without magnetic field, in the temperature regime around $T_c$ (~ 78K), for x=0.11.

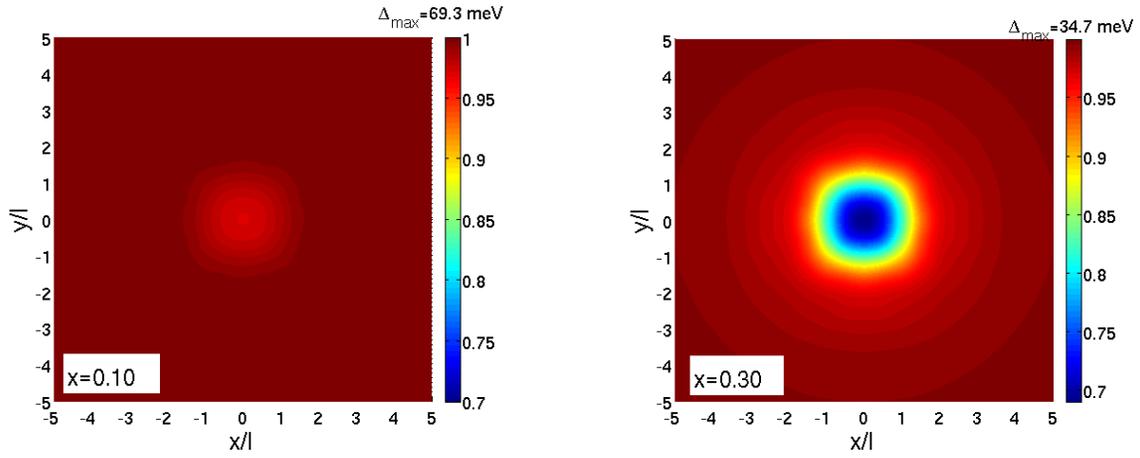

FIG. 9: Colour coded representation of the order parameter magnitude as a function of distance from the vortex core for a single vortex, at T=0 for two dopings, namely x=0.10 and x=0.30. The vortex core is at x=0,y=0, in the middle of a square plaquette, and distances are in units of the lattice constant.

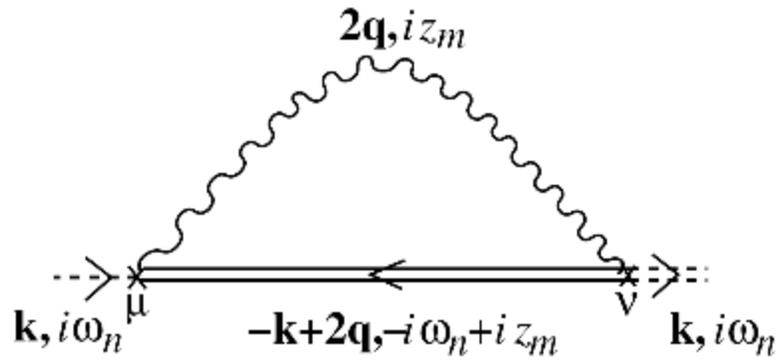

FIG. 10: Self energy diagram for an electron with momentum k and frequency (energy) $i\omega_n$ arising from the exchange of a Cooper pair of momentum q and frequency $iz_m$.

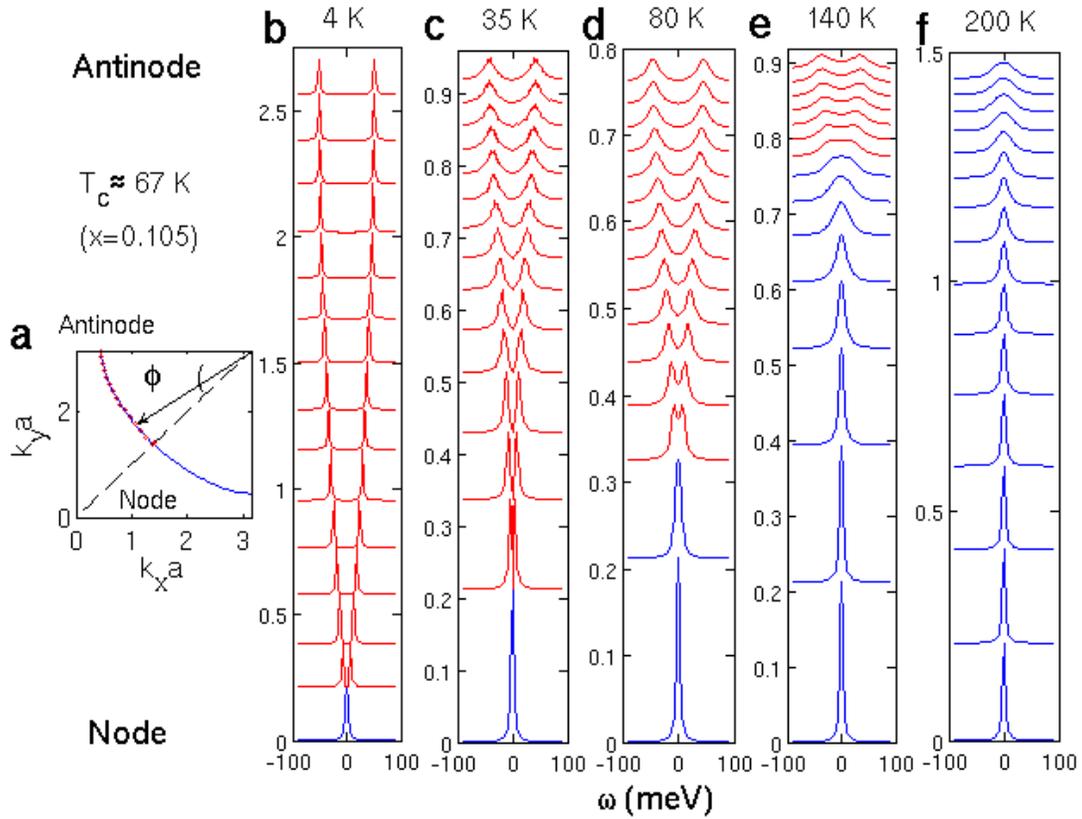

FIG. 11: Electron spectral density as a function of frequency (energy) ω in the Cooper pair fluctuation exchange approximation. The spectral density is shown for 15 uniformly increasing values of φ, namely the angle from the antinodal direction. (Panel (a) shows φ). The quantity is shown for the temperatures indicated in each panel (b)-(f). The spectral density curves of different values of φ are displaced with respect to each othervertically by the peak height of the one below it for clarity.

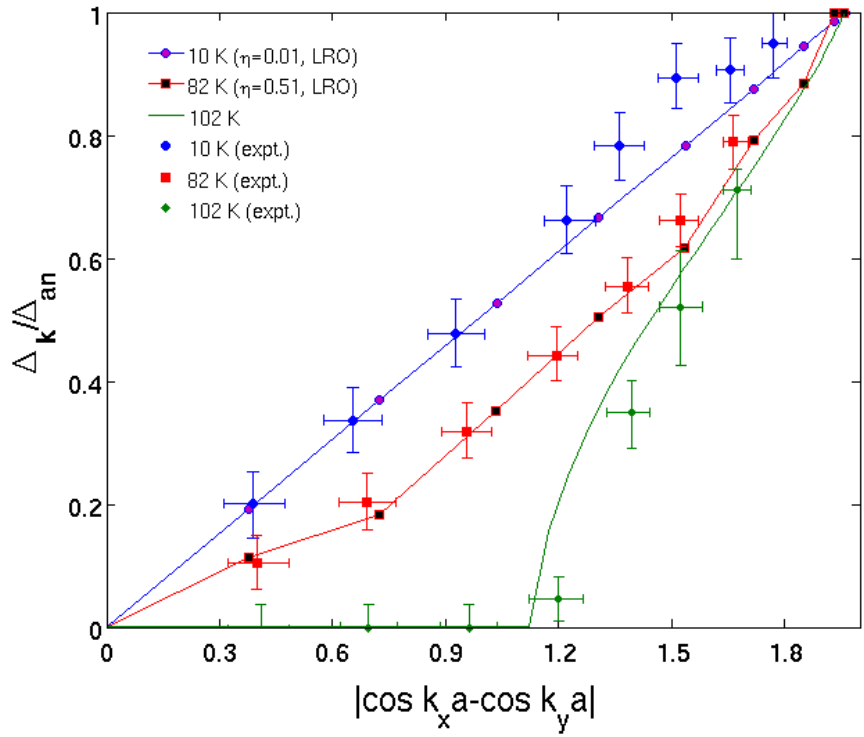

FIG. 12: 'Bending' of the gap $\Delta_k$ as a function of $|\cos(k_x a) - \cos(k_y a)|$. The full curves show our calculated values, the gap being obtained from the location of the spectral function peak; the experimental values are shown by various points.

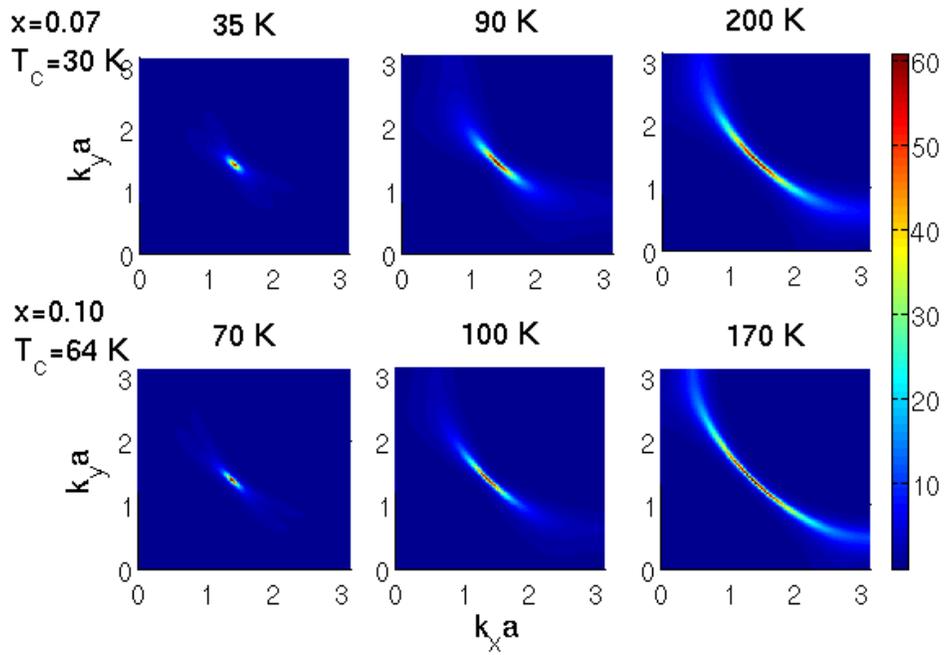

FIG. 13: Colour coded plot of spectral density for Fermi surface quasiparticles, for doping x=0.07 ( $T_c$ =30K) at the different temperatures shown. The 'Fermi arc' and its temperature dependence are quite strikingly visible.

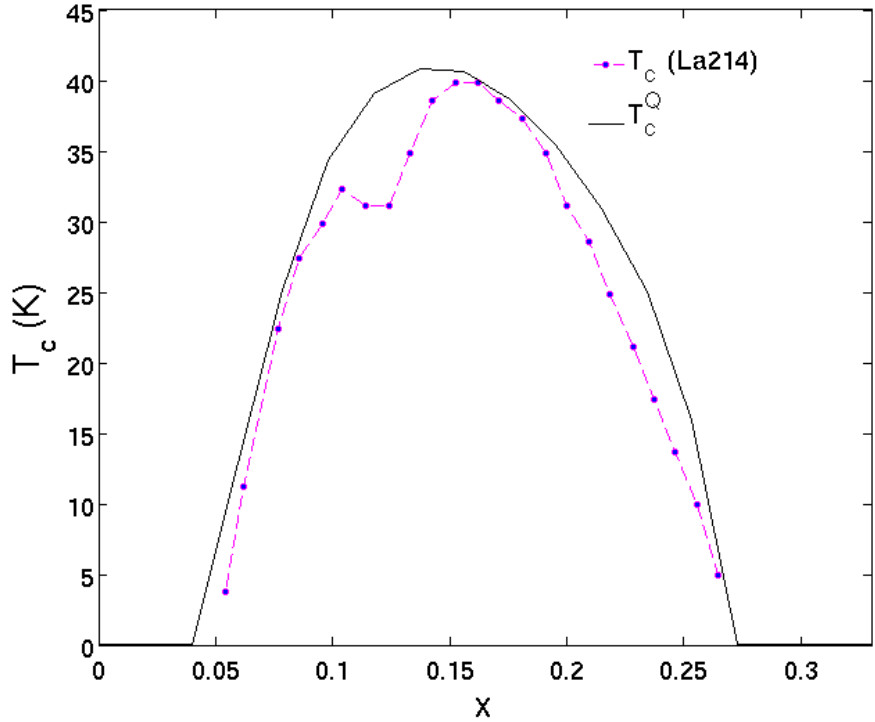

FIG. 14: Effect of quantum phase fluctuations on the superconducting transition temperature $T_c$, in a simple nearest neighbour model for the charging energy or the quantum phase fluctuation term. The continuous curve shows the calculated curve, compared to the observed $T_c$ for LSCO. The two have been matched to have the same $T_c$ at optimum doping x~0.19. We notice that due to quantum fluctuations, $T_c$ vanishes at nonzero x ~ 0.04 in the underdoped regime, rather than at x=0.00 as for classical phase fluctuations. In the overdoped regime, $T_c$ vanishes at x~0.28 rather than at about 0.35 as for classical fluctuations. The $T_c(x)$ curve is more symmetrical.